\documentclass[review]{elsarticle}

\usepackage{lineno,hyperref}
\modulolinenumbers[5]

\usepackage{bm}
\usepackage{graphicx}
\usepackage{amsmath,amsthm,amssymb,amscd}
\usepackage{epsfig}
\usepackage{color}
\usepackage{hyperref}
\usepackage{textcomp}

\journal{Neural Networks}

\biboptions{authoryear}

\begin{document}

\begin{frontmatter}

\title{Effect of dilution in asymmetric recurrent neural networks\tnoteref{t1,t2,t3}}
\tnotetext[t1]{This is the author version of the work.
This is the accepted manuscript version of the article published in Neural Network Journal.
Published journal article link: \href{https://doi.org/10.1016/j.neunet.2018.04.003}{https://doi.org/10.1016/j.neunet.2018.04.003}.}
\tnotetext[t2]{
Please cite this article as: Folli, V., Gosti, G., Leonetti, M., Ruocco, G., Effect of dilution in
asymmetric recurrent neural networks. \emph{Neural Networks} (2018),
\href{https://doi.org/10.1016/j.neunet.2018.04.003}{https://doi.org/10.1016/j.neunet.2018.04.003}.
}
\tnotetext[t3]{
\textcopyright 2018. This manuscript version is made available under the CC-BY 4.0 license.
\href{https://creativecommons.org/licenses/by/4.0/}{http://creativecommons.org/licenses/by-nc-nd/4.0/}.}

\author[mymainaddress]{Viola Folli}
\ead{viola.folli@iit.it}

\author[mymainaddress]{Giorgio Gosti\corref{mycorrespondingauthor}}
\cortext[mycorrespondingauthor]{Corresponding author}
\ead{giorgio.gosti@iit.it}

\author[mymainaddress,mythirdaddress]{Marco Leonetti}
\ead{marco.leonetti@roma1.infn.it}

\author[mymainaddress,mysecondaryaddress]{Giancarlo Ruocco}
\ead{Giancarlo.Ruocco@roma1.infn.it}

\address[mymainaddress]{Center for Life Nanoscience,\\
Istituto Italiano di Tecnologia,\\
Viale Regina Elena 291, 00161,  Rome, Italy}
\address[mysecondaryaddress]{Department of Physics,\\
Sapienza University of Rome,\\ 
Piazzale Aldo Moro 5, 00185, Rome, Italy}
\address[mythirdaddress]{CNR NANOTEC-Institute of Nanotechnology c/o Campus Ecotekne,\\
University of Salento, Via Monteroni, 73100 Lecce, Italy}

\begin{abstract}
We study with numerical simulation the 
possible limit behaviors
of synchronous discrete-time deterministic
recurrent neural networks composed of $N$ binary neurons 
as a function of a network's level of dilution and asymmetry.
The network dilution measures the fraction of neuron couples that are connected,
and the network asymmetry measures to what extent
the underlying connectivity matrix is asymmetric.
For each given neural network,
we study the dynamical evolution of all the
different initial conditions, thus characterizing the full dynamical 
landscape
without imposing any learning rule.
Because of the deterministic dynamics,
each trajectory converges to an attractor,
that can be either a fixed point or a limit cycle.
These attractors form the set of all the possible limit behaviors of the neural network.
For each network we then determine the convergence times, the limit cycles' length,
the number of attractors, and the sizes of the attractors' basin.
We show that there are two network structures that maximize the
number of possible limit behaviors.
The first optimal network structure is fully-connected and symmetric.
On the contrary, the second optimal network structure is highly sparse and 
asymmetric.
The latter optimal is similar to what observed in different 
biological neuronal circuits.
These observations lead us to hypothesize that 
independently from any given learning model,
an efficient and effective biologic network that stores
a number of limit behaviors close to its maximum capacity
tends to develop a connectivity structure
similar to one of the optimal networks we found.
\end{abstract}


\end{frontmatter}


\section{Introduction}
Recurrent neural networks are able to store stimuli-response associations,
and serve as a model of how
live neural networks store and recall behaviors as responses to given stimuli.
A discrete-time deterministic recurrent $N$ binary-neuron neural network 
is completely characterized by its $N^2$ edges, and 
its instantaneous state is defined by a neuron activation vector $\bm \sigma$,
which is a binary vector of size $N$.
In this paper, we consider a specific kind of recurrent neural network,
which is initialized, analogously to a Hopfield network, by assigning to the network's neurons 
an initial pattern which is the
the network stimulus or input.
The collection of all possible neuron activation vectors contains $2^N$ allowed vectors $\bm \sigma$,
these vectors can be partitioned in three categories:
steady states,
limit cycles, and transient states.
Steady states are neuron activation states that do not change in time,
and limit cycles are sequences of neuron activation vectors that repeat cyclically, with a period that we call cycle length.
From now on, we will consider a steady state
as a limit cycle of length 1.
A network, given any initial activation vector,
always evolves to a limit cycle, which for this reason we also refer to as attractor. 
In other words, a network associates a limit cycle to any initial neural activation state
that is given as an input.
For this reason, limit cycles can be considered as
behaviors
stored as responses to initial stimuli.
In the case of length 1 cycles,  limit behaviors are a
single activation state which in the case of Hopfield networks correspond to
the recollection of a memory. In the case of cycles with a length greater than 1,
stored limit behaviors are sequences of activation patterns which may correspond to
a stored dynamical sequence, such as the performance of a complex motor task, or a dynamic
sequence of static memories.
In principle, a recurrent neural network stores a certain number
of limit behaviors as vectors
from a $2^N$ set in a data structure defined by $N^2$ parameters. 
Furthermore, these vectors can be recovered in responses to input stimuli.
This clearly has intriguing analogies with content-addressable memory systems
capable of indexing large strings of bits \citep{Hopfield1982,Carpenter1989}.

In the past, recurrent neural network,
and specifically Hopfield neural networks
have been used to model memory storage and recall, though more recently
neurobiology models implemented recurrent neural networks
to describe brain activity in different cognitive tasks. \cite{Mante2013}
use recurrent neural networks to model the integration of context information in the
prefrontal cortex in discrimination tasks.
Similarly,
\cite{Carnevale2015} model
with recursive neural networks the premotor cortex modulation of its response criteria
in a detection task with temporal uncertainty. Furthermore recurrent neural networks
are used to model phoneme acquisition \citep{Kanda2009}, and language acquisition
\citep{Heinrich2018}.
Neuroscience proposes two fundamental
conceptual frameworks that enable recurrent neural networks
to store limit behaviors:
the connectionist hypotheses and the 
innate hypotheses. \cite{Hebb1949}
proposed the connectionist hypothesis,
which assumes that a neural network starts blank and 
forms new links or adjusts the existing ones each time it stores a new limit behaviors.
In this framework limit behaviors are stable
equilibria in the neural network dynamics.
A criticism of Hebbs' networks is that
as new limit behaviors are added
the corresponding generated stable states start interfering with the
stable states associated
with older limit behaviors. 
This limits the maximum storage capacity $C$,
which is defined as the maximum number of limit behaviors that can be stored.
Notice that this definition is different from the usual definition used in
associative memory networks, in which the storage capacity is
defined as the number of uniformly distributed random vectors 
that can be stored in an associative memory \citep{Hassoun1993,Hassoun}.
\cite{Amit1985a} shows that a Hebbian network has a storage capacity of $C=pN$ with $ p \approx 0.14$.
In contrast, innate network models assume that
limit behaviors are stored
using innate neural assemblies with a given connectivity. 
Among other innate memory models,
\cite{Perin2011} proposes that groups of pyramidal neurons in the rats' neocortex 
may be innate neuron assemblies that may only partially change their overall connectivity
structure.
Indeed, \cite{Perin2011} finds that these assemblies have similar connectivity proprieties among 
different animals, and argues that these assemblies serve
as building blocks for the formation of composite complex memories.

Whether we assume a connectionist or an innate network
scheme as our working framework,
we implicitly assume that
a recurrent neural network acts
as a content-addressing memory which given an input pattern (stimulus)
returns a limit behavior.
This limit behavior can be a recovered memory or a more complex neural sequence of neural activations
that may be integrated into a second neural network.
To understand how well a recurrent neural network
acts as a content-addressing memory,
the memory storage and retrieval literature
uses discrete-time recurrent neural networks
with McCulloch-Pitts neurons \citep{McCulloch1943}. 
Each discrete-time recurrent neural network,
which in this literature is sometimes referred to as Hopfield neural network, is characterized by its connectivity matrix $\mathbf J$, which schematically
represents the set of synapses and electrical junctions connecting couples of neurons.
Deterministic discrete-time synchronous recursive neural networks are deterministic discrete dynamical systems.
This implies three properties. First, each state in the neural network uniquely transits to another one.
Second, the reverse is not true, different states can evolve to the same state.
Third, each state belongs to a path that connects it to a stable activity pattern, \emph{i.e.} a limit cycle.
Given any initial neural state, or input, 
a discrete-time recurrent neural network
dynamically falls into an attractor.
In this framework, the attractor is
the retrieved limit behavior \citep{Hebb1949,Amit1985,McEliece1987,Folli2017,Gutfreund1988,Bastolla1998,Sompolinsky1988, Wainrib2013}. 
Finally, it is important to consider that 
a recurrent network
associates a limit behavior 
to each input from the set of all possible $N$-bit inputs, 
since the number of limit behaviors $C$ is such that $C<<2^N$, it performs a many-to-few mapping.
Recurrent neural network,
and in particular the Hopfield model \citep{Hopfield1982},
show how information can be stored via attractor states.
Indeed, there is some experimental support for discrete attractors in the rodents hippocampus cells' activity \citep{Pfeiffer2015},
and in monkey cells' activity during tasks \citep{Fuster1971,Miyashita1988}.

To understand how well and how many limit behaviors a fully developed neural network can store,
we explore how the structure properties of an arbitrary connectivity matrix $\mathbf J$
influences the attractor states of the network without imposing an a priori learning rules.
Given a connectivity matrix $\mathbf J$, 
to characterize
the network structure,
we define the network's asymmetry degree $\epsilon$, and dilution degree $\rho$.
The most understood properties on fixed discrete-time recurrent neural networks
regard fully-connected Hopfield neural networks.
Fully-connected recurrent Hopfield neural networks
are networks with dilution degree $\rho=0$, in which
any couple of neurons is connected by two axons one in each direction. 
In contrast, we define diluted recurrent Hopfield neural networks as networks with $\rho>0$,
in which only a subset of all neurons couples are connected.
The existing recurrent Hopfield neural network
literature mostly discusses symmetric neural networks in which the weights of the 
two axons connecting neurons $i$ and $j$ in both directions are the same,
and in only few cases researchers investigate 
asymmetric neural networks $\epsilon>0$, in which the weights are no longer equal.
Furthermore, most of the recurrent Hopfield neural network literature 
which studies the effect of asymmetry
assumes binary neurons with activations
state that can take values $-1$ and $+1$. 
Under these constraints, it is reported that symmetric fully-connected networks, $\epsilon=0$, have several attractors, all of which are formed
by cycles of length 1 and 2. As the network becomes less symmetric, $\epsilon>0$,
the attractors are composed of longer neural activation patterns.
Increasing asymmetry in a fully connected neural network introduces severe drawbacks.
Indeed, when the degree of asymmetry is increased above a certain threshold
a neural network is subject to a transition from an ordered phase to a ``chaotic" regime \citep{Bastolla1998,Gutfreund1988}.
In the chaotic regime, almost identical initial patterns can reach different attractors,
and the network is characterized by a high sensitivity to initial conditions.
Moreover, this chaotic regime causes exponentially longer recognition time,
where the recognition time is the average number of discrete transitions required to reach the corresponding attractor
from a generic point in its basin of attraction.
\cite{Toyoizumi2015} analyzes asymetric matrices $\epsilon=0$ with neuron activation profile $\{-1,1\}$,
and shows that under these conditions as the limit cycle lenght scales exponentialy with $N$,
the number of attractor scales linearly with $N$. 
It is important to point out that \cite{Bastolla1998}, \cite{Toyoizumi2015}, and \cite{Gutfreund1988}
discuss synchronous discrete-time recurrent neural networks with single neuron activation profile $\{-1,1\}$.
In this work, we show that these results carry over to neurons with activation profile $\{0,1\}$.
Neurons with activations states $\{0,1\}$ represent
a more accurate model of the single neuron
behavior that is observed as either firing or at rest,
and how these states determine the excitatory
or inhibitory interactions among the neurons in a live neural network.
From these evidence, we would argue that symmetric networks, $\epsilon=0$, are
the optimal limiting state for a developing neural network.
Nevertheless, most natural neural networks are asymmetric, $\epsilon>0$ \cite{Perin2011}.
Therefore, a puzzling contradiction emerges between the observed asymmetry 
in natural synaptic connections,
and the disastrous properties that emerge in fully connected neural network models
when the connectivity matrix becomes more asymmetric.
Furthermore, in almost all cases live neural networks are not fully connected,
$\rho=0$, but tend to be highly diluted $\rho \sim 0.9$,
which implies that not all but just a fraction of neuron couples are connected in the matrix $\mathbf J$.
For this reason, this paper analyzes not only how the storage properties of neural networks change as we change 
the asymmetry degree of fully connected connectivity matrices $\mathbf J$, $\epsilon \in [0,1]$ and $\rho=0$.
It also studies how the storage properties change as
asymmetric connectivity matrices $\mathbf J$ are diluted, $\epsilon = 1$ and $\rho \in [0,1]$.

Because under certain connectivity conditions we get recurrent neural networks
characterized by a minority of attractors with
huge attraction basins and a majority of attractors with small attraction basins,
almost the entire state space of the neural network is absorbed by few attractors.
Consequently, we decided to map the corresponding attractor pattern for each
initial network condition.
This allows us to explore the entire landscape
generated by the complete mapping of all the attractors' basins for different connectivity matrices $\mathbf J$
without losing the attractors with small basins.
In addition, we define a way to sample from the space of all connectivity matrices
a subset of connectivity matrices $\mathbf J$
with a given asymmetry degree $\epsilon$, and sparsity degree $\rho$.
Thus, we can chart how the attractor landscape properties change 
sampling connectivity matrices $\mathbf J$ for different
values of asymmetry degree $\epsilon$, and sparsity degree $\rho$. 
Unfortunately, the computation of the complete transition landscape 
over all the $2^N$ initial states 
is computationally feasible
for neural networks up to a certain  $N$.
Nevertheless, given these limitations we find the 
non-trivial result that 
the symmetric/fully connected region is not the only region with
optimal storage capacity, but a second optimal region, the
fully-asymmetric/high-dilution region, 
exists as well.
The scaling trends observed in simulations lead us to hypothesize that the results are valid also for values of $N$ larger than
the numerical computations we were able to perform.
It is surprising to notice that both regions exhibit a similar scaling of the storage capacity
even if they have very different connectivity.

Additionally, the values of sparsity and asymmetry, which optimize the number of limit behaviors in discrete-time recurrent neural networks,
are found to be remarkably similar to those found in several regions of the mammalian brain that share crucial roles in memory processes.
Indeed, the neocortex and the CA3 region of the hippocampus
have been proposed as regions responsible for memory storage \citep{Rolls2012},
and are clearly diluted \citep{Witter2010}.
The probability of connection between two neocortical pyramidal cells is in the order of $10\%$,
and the probability of connection between two hippocampal CA3
neurons is nearly $4\%$.
While the symmetric/fully optimal connected networks
are consistent with the observed live neural networks,
the same networks
contradict a standard interpretation of Hebbs' Learning, which 
are noted to tend to the asymmetric/dilute optimal networks.
Thus, we are left with the question of why biological networks are found in the
fully-asymmetric/high-dilution region. There may be several
possible explanations that beg for further investigation.
One possibility 
is that fully connected networks are more costly
to develop and maintain in comparison with diluted networks, especially for larger values of $N$,
because for large $N$ there could be important constraints that limit
the development of all possible couples of neurons.

The role of dilution in natural neural networks was ignored in past research
with the exception of only a couple cases.
Theoretical mean field approaches on diluted recurrent neural networks
showed weakly correlated firing patterns similar to the patterns observed 
in the brain cortex \citep{vanVreeswijk1998, Monteforte2012}.
\cite{Kim2017} used simulation to show how, in contrast to fully connected
recurrent networks, scale-free networks have an increased
number of final behaviors
with the side effect of increased errors.
\cite{Brunel2016} is an other work that comes to the conclusion that diluted network
have optimal storage capacity.
This later work investigates the connectivity structure 
of recurrent neural networks with excitatory synaptic connectivity which are close to their storage limit
given a certain degree of retrieval robustness
independently from the learning rule.
To this aim \cite {Brunel2016} assume that the inhibitory interactions
are fully-connected and finds the excitatory matrices that store the greatest number of attractors
both analytically with the cavity method and numerically with perceptron learning.
Then it investigates the statistical properties of
these optimal matrices in large networks.
Brunel found that, the optimal excitatory connectivity matrix is diluted,
it has several zero-weight synapses, and that the number of neuron couples with
reciprocal connections are greater than in a random network, and that their weight is larger than the average
connection weight.
Similarly, the approach of this paper is independent of the learning rule, and
it searches for the properties of the networks that have optimal storage capacity.
Nevertheless, instead of calculating a neural network that can store the maximum number of attractors,
it assumes that a network which has reached its storing capacity limit has
certain characteristic optimal structures. Thus to find this optimal connectivity structures
we can sample sets of connectivity matrices $\mathbf J$ 
with given values of asymmetry $\epsilon$ and dilution $\rho$.
Each of these sets will have a characteristic 
asymmetry degree $\epsilon$ and dilution degree $\rho$
and be composed of connectivity matrices $\mathbf J$ with certain memory storage properties.
From these sets, we can find the values of asymmetry and dilution
which characterize the connectivity matrices $\mathbf J$ with optimal memory storage properties.
Discrete-time recurrent neural networks
are minimal models designed to capture the fundamental nature of neural networks,
and
more realistic models have been proposed 
\citep{Gerstner1998,Maass1999, Koch1998, Galves2013}.
Unfortunately, these models have a necessary additional computational  
cost which would not allow for such extensive exploration
of the attractor landscape associated to connectivity matrices $\mathbf J$
at different degrees of dilution and asymmetry.

\noindent \section{Discrete-Time Recurrent Neural Networks}

We consider a network of $N$ binary neurons interacting via a connectivity matrix $\mathbf J$,
with matrix elements $J_{ij}$ for $i,j=1,...,N$.
The matrix element $J_{ij}$ represents the strength of the connection between
the pre-synaptic neuron $j$ and the post-synaptic neuron $i$.
The state of each neuron is represented by a binary state variable $\sigma_i \in \{0,1 \}$,
where $i=1,\dots,N$. A neuron $\sigma_i$ takes values either $0$ or $1$ if it is respectively at rest (inactive) or firing
(active).

\subsection{Dynamics}
In this section, we introduce a synchronous discrete-time recurrent neural network model with McCulloch-Pitts \citep{McCulloch1943} neurons.
At each time step, all neurons are updated synchronously according
to the discrete-time recurrent neural network evolution rule:
\begin{equation}
\label{eq1}
\sigma_i(t+1)=\theta\Big(\sum\limits_{j=1}^N J_{ij}\sigma_j(t)-\eta_i\Big),
\end{equation}
where $\theta(x)$ is the Heaviside step function ($\theta(x) = 1$ for $x \ge 0$, and $\theta(x) = 0$ otherwise). 
At the next step $t+1$, the neuron $i$ fires, $\sigma_i(t+1)=1$,
if the summation of its synaptic inputs is above a threshold $\eta_i$, otherwise the neuron is inactive, $\sigma_i(t+1)=0$. In the results discussed in this paper we set $\eta_i = 0$
for all neurons.
The vector $\bm \sigma(t) = ( \sigma_1(t),\sigma_2(t),\dots,\sigma_N(t) )$
represents the activation profile of all neurons at time $t$.
This activation function emulates the \textit{all-or-none} principle in neuronal activation potentials.
The input summation is performed without any scaling on $N$ since the threshold is equal to zero.
The scaling factor does not influence the network output that is
indeed determined only by the sign of the linear summation.
In principle neurons update asynchronously, but
\cite{Gutfreund1988}, and \cite{Nutzel1991} observed that
synchronous updating neurons are qualitatively equivalent to asynchronous updating neurons
when considering long time scales.
Because the objective is to map the full attractor landscape,
we have to explore all the possible initial conditions.
For this reason, it is fundamental to consider a simplified scheme,
and we chose to use a parallel synchronous evolution rule to simplify and speed-up the numerical simulations.

\subsection{Content-Addressing Memory Model}

A synchronous discrete-time recurrent neural network is a deterministic dynamical system
defined on a finite state space. Given any initial
neural activation profile $\bm \sigma(0)$ the discrete-time recurrent neural network 
deterministically evolves in time until it converges to an attractor, which can be
composed of a limit cycle composed of a certain number of
neural activation states.
When a query is submitted to a content-addressing memory
we input a search entry 
and the device returns an address to matching
stored data. 
A discrete-time recurrent neural network
works analogously to a content-addressing memory.
If the initial neural activation profile $\bm \sigma(0)$
is set equal to an external stimulus, or search entry in the content-addressing memory analogy,
then the neural network
dynamically converges to an attractor, which represents the retrieved responce.
Given this analogy, the limit behavior storage capacity $C$ 
of a discrete-time recurrent neural network
is the number of attractors that are stored in the network.

According to Hebb's rule, memories are added to a Hopfield network
by adding to the neural connectivity matrix $\mathbf J$ dyadics of the  
form $\bm \sigma^T \bm \sigma$ \citep{Hebb1949}.
These dyadics generate attractors formed by a length 1 cycle.
Thus, in Hebbian learning a memory is a limit behavior composed of
a single neural activation profile $\bm \sigma$.
\cite{Amit1985a} demonstrated that there is an upper limit to the storage capacity of Hebbian learning networks,
and this limit is set to $pN$ with $p \approx 0.14$. 
This storage capacity upper limit
emerges because each time we add a new dyadic term
a new attractor is introduced in the Hopfield network dynamics,
and at a certain point the new dyadic terms 
associated with the latest observed attractor interferes with the basin of the old attractors.

In this paper, we use an approach that is independent from the specific learning rule.
We assume that, independently from the learning rule,
when a discrete-time recurrent neural network
stores a number of limit behaviors close to its storing capacity
then it approaches characteristic optimal connectivity structures.
Thus, instead of packing as many memories as we can
in a discrete-time recurrent neural network,
and then studying the emergent network structure. We 
use the total number of attractors to count the number of the network's
limit behavior storage capacity.
Clearly, some of these attractors may be spurious associations
that were not explicitly learned.
Consequently, we search for connectivity structures
that form the recurrent neural network 
with the maximum number
of attractors. 
More precisely, we study how many attractors are present in a given arbitrary neural network with 
a certain connectivity matrix $\mathbf J$ characterized by a specific degree of asymmetry and dilution.
Finally, 
we can study how a
recurrent neural network 
with an optimal connectivity matrix
reduces the complexity of the $N$-dimensional initial problem
by clustering the input data in a certain number $C$ of attraction basins. 

\subsection{Synaptic Matrix}\label{sec:mat}

$\mathbf J$ is the neural network's connectivity matrix 
with matrix elements $J_{ij}$. To analyze the relation between the full attractor landscape and certain global
properties of $\mathbf J$, we defined a way to generate a random connectivity matrix $\mathbf J$
given two fundamental network parameters, respectively the asymmetry coupling degree $\epsilon$, and the sparsity
parameter $\rho$. This section first describes how to generate a random connectivity matrix $\mathbf J$
with an arbitrary asymmetry degree $\epsilon$, and fixed sparsity parameter $\rho = 1$, then it describes
how to generate a random connectivity matrix $\mathbf J$ with arbitrary values of $\epsilon$ and $\rho$.

The elements of a random connectivity matrix $J_{ij}$
with an arbitrary asymmetry degree $\epsilon$ and fixed sparsity parameter $\rho = 1$
are generated
as the convex sum of the matrix-elements from
a symmetric random matrix $S_{ij}$,
and the matrix-elements from a antisymmetric random matrix $A_{ij}$:
\begin{equation}
\label{eq2}
J_{ij}=\left(1-\frac{\epsilon}{2}\right)S_{ij}+\frac{\epsilon}{2}A_{ij}.
\end{equation}
$S_{ij}$ and $A_{ij}$ are generated with a three-step process.
First, the lower diagonal elements of
$S_{ij}$ and $A_{ij}$ are randomly drawn from a uniform distribution with
greater-than-zero probability in the closed interval $\left[-1,+1 \right]$.
${\cal{P}}_0(x)=\frac{1}{2}\theta(1-x^2)$ is the distribution from which both
the lower diagonal elements of $S_{ij}$ and $A_{ij}$ are drawn,
${\cal{P}}_0(S_{ij})$ and ${\cal{P}}_0(A_{ij})$.
Second, the upper diagonal elements are set:  
$S_{ji}$=$S_{ij}$ and $A_{ji}$=$-A_{ij}$. Lastly, the diagonal elements are set to zero:  
$S_{ii}=0$ and $A_{ii}=0$. 
$J_{ij}=0$ corresponds to no-connection between $i$ and $j$,
and an $J_{ij}$ element greater or smaller than zero corresponds respectively to an excitatory 
or an inhibitory connection.
For $\epsilon$=0, neurons interact symmetrically with each other; at $\epsilon$=1, $J_{ij}$ is fully-asymmetric.
For $\epsilon$ ranging from $1$ to $2$, the strength of antisymmetric couplings increases.
In this work, we restrict the analysis for $\epsilon\in[0,1]$.
We set the diagonal elements to zero, $J_{ii}$=0 which implies no autapses (the term autapse
indicates a synapse connecting a neuron onto itself).

In order to generate a random connectivity matrix $J_{ij}$
with an arbitrary asymmetry degree $\epsilon$ and arbitrary sparsity parameter $\rho $,
we first generate $S_{ij}$ and $A_{ij}$, and then use Eq. \ref{eq2} to form $J_{ij}$.
To generate $S_{ij}$ and $A_{ij}$ we draw the elements of the respective lower diagonal elements according to the distribution
\begin{equation}
\label{eq4}
{\cal{P}}(x) = (1-\rho){\cal{P}}_0(x)+\rho\delta(x),
\end{equation}
where $x$ takes the place of $S_{ij}$ and $A_{ij}$, and $\delta(x)$ is the Dirac function.
${\cal{P}}(x)$ ensures that the rate of zero-valued elements is determined by the sparsity parameter $\rho$.
Indeed, a random matrix according to this distribution can be generated
by drawing $S_{ij}$ and $A_{ij}$ elements
from ${\cal{P}}_0(x)$,
and then setting each element with probability $\rho$ to zero.
In (\ref{eq2}), the left term determines the ratio and distribution of non-null elements,
and the right term determines the ratio of null elements.
Then, the upper diagonal elements are set to  
$S_{ji}$=$S_{ij}$ and $A_{ji}$=$-A_{ij}$, and the diagonal elements are set to zero. 
Finally, we construct $J_{ij}$ accordingly to (\ref{eq2}).
This procedure preserves the symmetry and the role of $\epsilon$. Indeed, while $\epsilon=0$ represents a symmetric matrix with a fraction of zeros on average equal to $\rho+1/N$, $\epsilon=1$ indicates a generic, asymmetric, matrix with a number of zeros
on average equal to $\rho^2 +1/N $, which
on average is approximately equal to $2\rho - 1 +1/N$ for $\rho$ close to one.
$\rho$ determines how many elements of the matrix are null, and thus the dilution
of the neural network connectivity. For $\rho=0$ we have a fully-connected neural network, in which any two
neurons share a synaptic connection in both directions. In the limit for $\rho$ that gets closer to $1$,
we get a neural network with constantly fewer neural connections, and for $\rho=1$ we have a network
completely lacking any synapses.

\subsection{Neural Network Dynamics}

\begin{figure}[h!]
\includegraphics[width=\columnwidth]{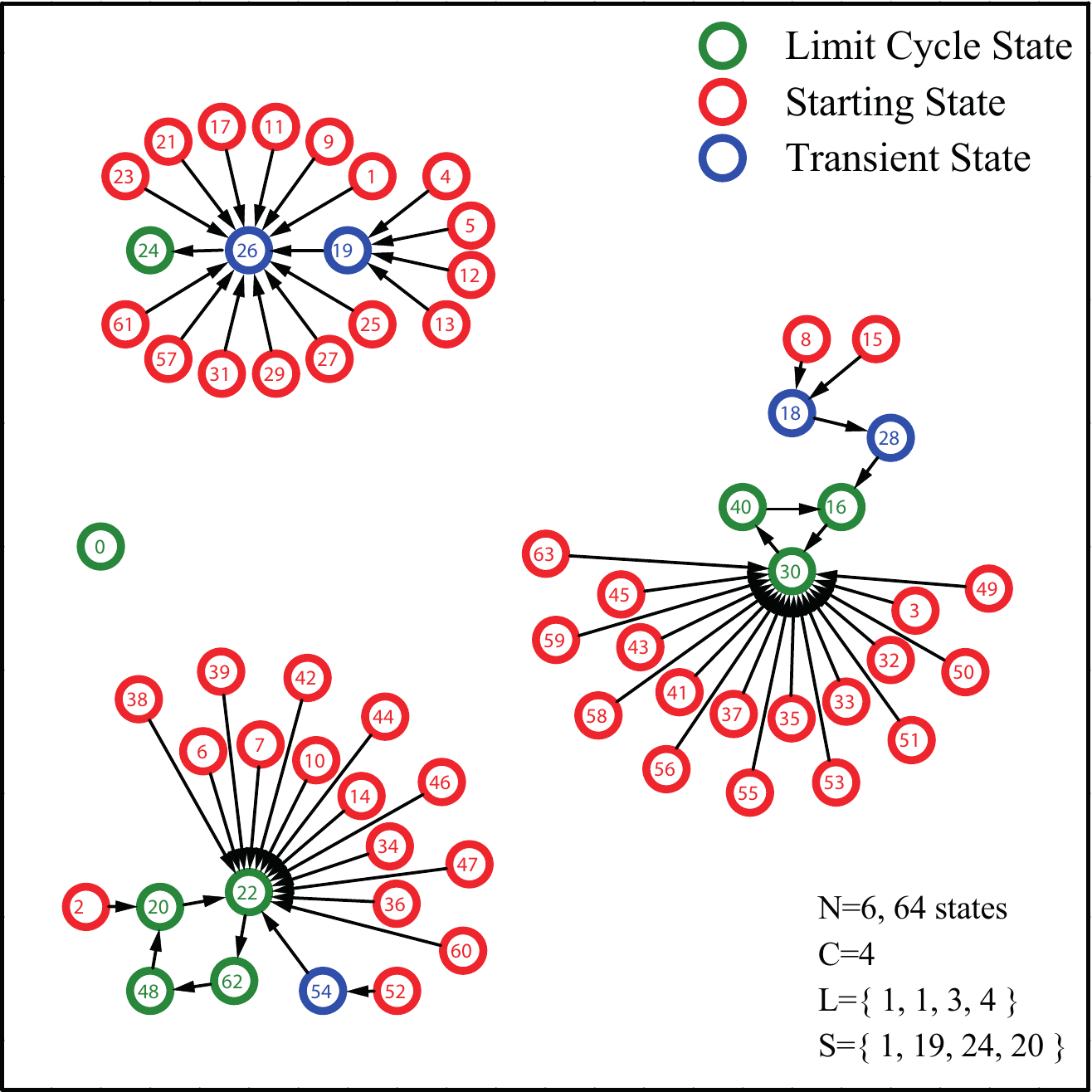}
\caption{(Color online) Example of weakly connected components of the transition graph $\mathcal{G}$ derived from a given 
connectivity matrix $\mathbf J$. The $2^N$ numbered circles represents the neural activation profiles $\bm \sigma $.
In the present case, $N$=6, and the connectivity matrix $\mathbf J$ is such that $\epsilon$=0 and $\rho$=0.
There are four weakly connected components ($C=4$), each one containing one attractor,
two of them are fixed points, $L_1=L_2=1$, the other are cycles with length $L_3=$3 and $L_4=$4.
The size of their attraction basins are $S_1=1$, $S_2=19$, $S_3=24$, and $S_4=20$ respectively.
Green circles indicate the points belonging to attractors, while blue and red circles indicate 
respectively transient and starting points.} \label{fig1} 
\end{figure}

Generated a connectivity matrix for a given pair $(\epsilon, \rho)$, we evolve all $2^N$ initial conditions.
Being the configuration space finite and the dynamics deterministic,
each state $\bm{\sigma}(t)$ uniquely transits to a state $\bm \sigma(t+1)$.
Necessarily, after a transient time, the system relaxes on a limit cycle. 
Given a certain $\mathbf J$, let $\mathcal{G} = (\bm \sigma, T)$ be a directed graph with nodes given by the neurons profiles $\bm \sigma$,
and edges $T$ defined as directed couples such that $(\bm \sigma,\bm \sigma') \in T$ if $\bm \sigma$
transits to $\bm \sigma'$. All nodes in $\mathcal{G}$
have one and only one outgoing edge.
Let $\mathcal{U}$ be the undirected graph associated to the directed graph $\mathcal{G}$, such that
for each $(a,b) \in \mathcal{G}$ there is an undirected couple $\{a,b\} \in \mathcal{U}$.
Furthermore, let a weakly connected component of a graph $\mathcal{G}$ be a maximal set of nodes $W$ such that
for each couple of nodes $(a,b) \in W$ there is a path that follows the 
edges in $\mathcal{U}$
and connects $a$ to be $b$.
Since in discrete-time recursive neural networks each initial neuron profile of all the possible $2^N$
patterns arrives at the corresponding attractor,
$\mathcal{G}$ is partitioned in a finite set of $C$ weakly connected components
each leading to an attractor \citep{Diestel2010}.
An example of this decomposition is reported in Fig. (\ref{fig1}) for a simple $N$=6 case.
This specific network maps $64$ possible states into four attractors: two limit cycles with respectively $L=3,4$ and two fixed points.

For each weakly connected component $k$, where $k=1\dots C$, and $C$ is the number of attractors,
we measure the following observable variables:
$(i)$ the cycle length $L_k$ is the number of states in its attractor, if $k$ is a fixed point $L_k = 1$
or if it is a limit cycles $L_k > 1$;
$(ii)$ the basin of attraction size $S_k$ corresponds to the number of states 
in the $k$-th weakly connected component; $(iii)$ the average distance $D_k$ between a generic state in the 
$k$-th weakly connected component and the corresponding attractor (see table~\ref{tab1}).

\begin{table}[h]
\centering
\caption{Measured quantities.}
\label{tab1}
\begin{tabular}{|c|c|c|}
\hline {\bf Quantity} & {\bf Description}\\
\hline $C$ & Number of attractors per matrix\\
\hline $L_k$ & Attractor length \\
\hline $S_k$ & Basin of attraction size \\
\hline $D_k$ & Average distance from the attractor \\
\hline
\end{tabular}
\end{table}
We generate $R$ replicas of the connectivity matrix $\mathbf J$ with given $\epsilon$ and $\rho$ (up to 100000),
and average the values of $C$, $L_k$, $S_k$, and $D_k$.
Then we produce the histograms
${\cal{N}}(C|\epsilon,\rho)$, ${\cal{N}}(L_k|\epsilon,\rho)$, ${\cal{N}}(S_k|\epsilon,\rho)$, ${\cal{N}}(D_k|\epsilon,\rho)$.

\subsection{Large $N$ Limit}

All the quantities in table~\ref{tab1} depend on the network size $N$. 
Given a  network sizes $N$, we compute the values of these quantities from
the complete transition graph $\mathcal{G}$ with all the $2^N$ neural activation profiles.
Thus we measure the scaling of these quantities for different values of $N$.
Because a statistical analysis would be biased by the broad distribution
of the basins sizes,
we study the complete graph $\mathcal{G}$. 
Indeed, a few basins are large with size $2^{N-m}$ where $m$$\ll$$N$, and several basins are very small in size. 
Thus, a statistical analysis would highlight only the trajectories absorbed by the largest basins,
given that the probability of selecting the smaller basins is very small.
This would result in a under estimation of the number of independent limit cycles.
For this reason, 
it is mandatory to evolve all initial conditions to map the whole landscape of the equilibrium properties of the network. 
We can complete the analysis only up to a computationally feasible value of $N$,
which turns out to be $N \approx 22$. We have this computational bound
because the number of initial conditions scales like $2^N$,
and the time to evolve all these states and to find the 
associated attractors scales proportionally to $2^{2N}$.

In the numerical results presented in the next section
we show that the scaling proprieties of $N$
are mostly preserved across the explored range.
Moreover, the functional form and the eventual exponents
or multiplicative factors
depend on the network parameters $(\epsilon, \rho)$.
A reasonable question is whether or not the scaling exponents derived from these limited $N$ values
can be assumed to be valid in the thermodynamic limit, for large $N$.
To ensure that there are no major changes as $N$ diverges,
we have performed a preliminary investigation of the number of cycles of length 1,
${\cal{N}}_1$ for $\rho$=0 and $\epsilon$ in the range [0,1].
\cite{Tanaka1980} presents a theoretical prediction
for this quantity, ${\cal{N}}_1$=$e^{\gamma(\epsilon) N}$,
with known $\gamma(\epsilon)$
in the infinite size $N$ limit.
To test, if the scaling exponent numerically obtained from our simulations with $N<18$ matches the theoretical infinite size exponent,
we compared the estimated $\gamma(\epsilon)$ value, $\hat\gamma(\epsilon)$, from our simulations, 
with the theoretical $\gamma(\epsilon)$ found in \cite{Tanaka1980}.
We found that the two exponents agree within one part per thousand.
This result, though not conclusive, gives us confidence that the
finite size estimates of the scaling exponents can be plausible approximations in the large $N$ limit,
both in the dense and diluted cases.

Summarizing, this paper focuses on limited size neural networks, this gives us the possibility
to exhaustively explore the whole equilibrium landscape
and map its changes as symmetry and dilution are varied.
A vast amount of literature has been concerned with the analysis of the statistical properties of large neural networks
in the thermodynamic limit, using mainly mean-field theories,
but little is 
known about diluted networks.
The future objective is to apply the cavity method \citep{Mezard1987}, or the replica method \citep{Gardner1988},
to analytically determine the number of attractors for diluted connectivity matrices
with large $N$. Thus, we will be able to compare these analytical results with the computational results presented here.

\section{Results}
In this section, we numerically investigate the attractors' landscape 
of a discrete-time recurrent neural network for different values of the network size $N$,
the asymmetry degree ($\epsilon$) and the synaptic connectivity dilution ($\rho$). 
We report numerical results obtained averaging over $R\sim10^4$ replica matrices.
We start characterizing fully-connected matrices with a specific degree of asymmetry ($\rho=0$, $\epsilon$).
This class of matrices was already investigated analytically, and
numerically sampling over the initial conditions, and assuming neuron activation state $-1$ and $1$.
Unfortunately, as mentioned earlier
sampling on the initial conditions may lead to biased results that miss smaller attractors.
Furthermore, we use the more realistic model with neuron activation state $0$ and $1$.
Next, we characterize statistically high-dilution/asymmetric cases,
a region that so far has been unexplored.

\subsection{Fully Connected Network: from Symmetric to Asymmetric Networks}

\begin{figure}[h!]
\begin{center}
\includegraphics[width=0.95\columnwidth]{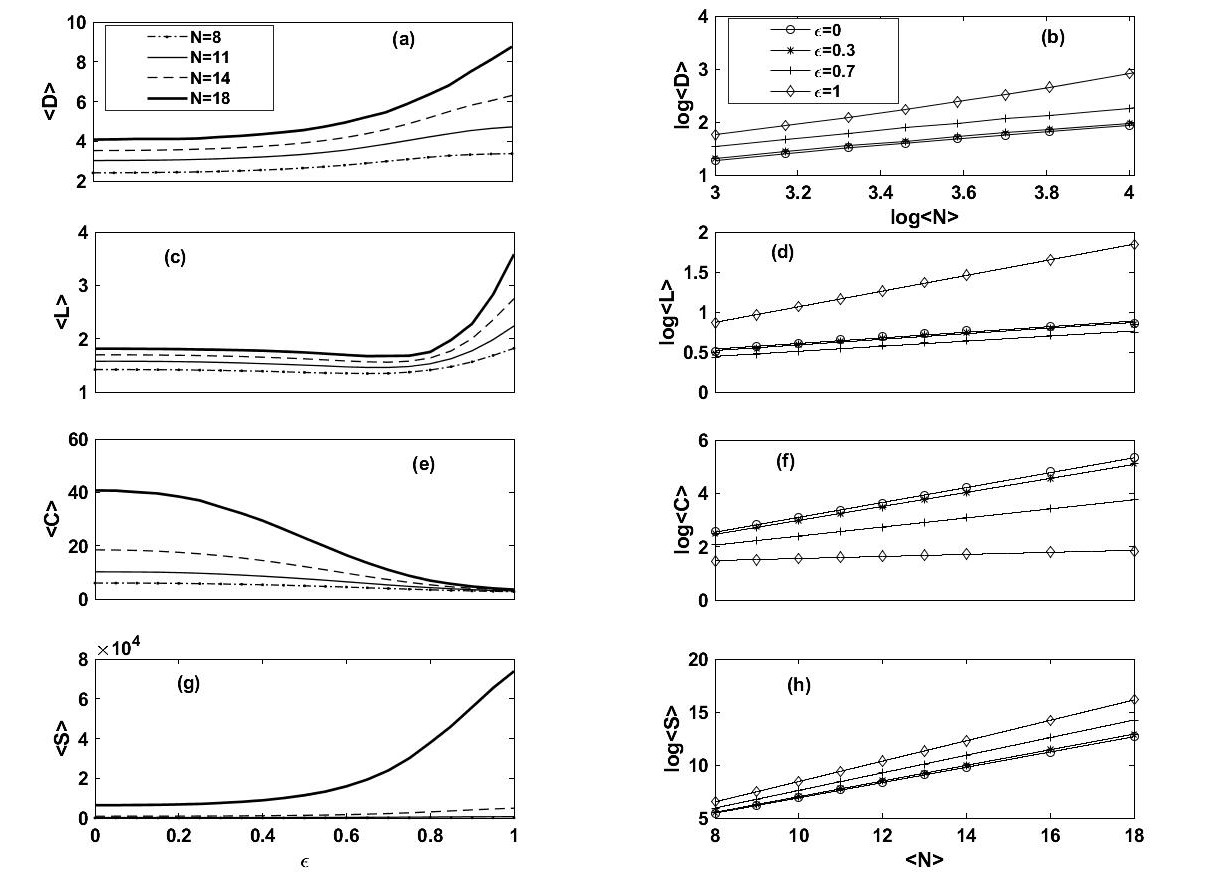}
\caption{The panels on the left column report 
the averaged quantities in Table~\ref{tab1} as a function of the asymmetry degree $\epsilon$ for $N=8,11,14,18$.
The panels on the right column display the corresponding scaling laws with the system size $N$
for $\epsilon=0, 0.3, 0.7, 1$.
(\textbf{a}) Increasing the asymmetry slows down the convergence time $\langle D \rangle$ to the attractor,
the corresponding scaling law (\textbf{b}) is polynomial with $N$.
(\textbf{c}) The asymmetry produces the emergence of limit cycles with length $L>2$,
at $\epsilon\sim0.75$ the averaged $\langle L \rangle$ increases exponentially $\sim 2^{\gamma N}$ with exponent $\gamma=0.1$ (\textbf{d}),
which indicates exponentially long limit cycles. (\textbf{e}-\textbf{f}) Increasing asymmetry in fully connected
neural networks leads to a drastic reduction of the number of attractors
which develop larger basins (\textbf{g}-\textbf{h}). This
implies a loss of computational ability,
because for larger $\epsilon$ the network becomes unable to cluster the input data in sub-domains.} 
\label{fig2}
\end{center}
\end{figure}

In figure~\ref{fig2}, we report all the quantities listed in table~\ref{tab1} measured 
for varying $\epsilon$ with $\rho=0$, and the corresponding scaling laws.
Here, and in the following, $\langle X \rangle$ indicates an average on $X$
taken over all the attractors, $k$, and all $R$ replicas of the connectivity matrix $\mathbf J$,
where $X$ is a quantity from Table~\ref{tab1}. 
As it is shown in (\textbf{a}), increasing asymmetry slows down the time required to convergence on the attractor. 
Thus, as asymmetry increases the stimulus-response time increases.
Plot (\textbf{b}) shows how $ \langle D \rangle $ scales with $N$
and that its trend is well reproduced by a polynomial law.
For the mean length of the limit cycles $\langle L \rangle$, we observe two distinct regimes, as clearly
presented in panel (\textbf{c}) where $\langle L \rangle$ shows a transition around $\epsilon=0.75$.
This transition for discrete-time recurrent neural networks 
with neuron activation states $\sigma_i \in \{0,+1\}$ is analogous to the
transition observed in recurrent neural networks
with $\sigma_i \in \{-1,+1\}$ \citep{Gutfreund1988}.
Correspondingly, the scaling law (\textbf{d}) displays a rather sharp transition from an ordered phase where
\begin{equation}
\langle L\rangle\sim N^{\gamma_{p}}
\end{equation}
to a chaotic regime, in which the mean length of limit cycles increases exponentially with the system size $N$:
\begin{equation}
\langle L\rangle\sim 2^{\gamma_{e} N}.
\end{equation}
An exponentially long limit cycle length means that for large $N$, the network displays
longer attractors with seemingly
aperiodic behavior (the network dynamics appears to be chaotic).

In panels $e$ and $g$, we report the mean values of the number of attractors 
per matrix, $\langle C \rangle$, and of the basins of attraction size $\langle  S \rangle $
for fully-connected matrices, $\rho = 0$.
It is clearly evident how the network loses the capacity to cluster the input data in sub-domains as $\epsilon$ is increased: the number of attractors
$\langle C \rangle$ is drastically reduced and correspondingly the dimension of the basins increases.
The state space tends to be subdivided into a small number of large regions, each with its attractor,
and the network is no longer able to distinguish different inputs. 
The average number of attractors scales exponentially for $\epsilon=0$
(panel $f$).
For fully-connected matrices, $\rho = 0$, and $\epsilon=0$, $\langle C\rangle \sim 2^{\gamma_{e} N}$ with $\gamma_{e}=0.28$.
For $\rho = 0$, and $\epsilon=1$, the number of attractors scales polynomially, as expected from \cite{Toyoizumi2015} analysis. In $\langle C \rangle$, there is probably 
a transition as for the attractor length $\langle L \rangle$. 
Panel ($h$) shows the scaling of $\langle S\rangle$.
For fully-connected matrices, $\rho = 0$, $\langle S\rangle$ has an inverse behavior compared to $\langle C \rangle$,
$\langle S\rangle$ scales exponentially for $\epsilon=1$, and it scales polynomially for $\epsilon=0$.
Increasing asymmetry causes the network to develop a glassy behavior,
the network is greatly slowed down and unable to quickly recognize the attractor of a given input. 
The higher the asymmetry degree, the higher the probability that the system admits only one attractor with a basin $2^N$
states large,
the network loses its capacity to deal with ``complexity".
In conclusion, although asymmetry in synaptic connection is a property of biological networks, asymmetry in
dense networks is not
sufficient to realize a network
capable of storing and retrieving stimulus response associations. We now investigate the
interplay between asymmetry and sparsity and how the dilution helps asymmetric networks
to store more memories.

\subsection{From Fully-Connected to Diluted Asymmetric Networks}

\begin{figure}[h!]
\includegraphics[width=\columnwidth]{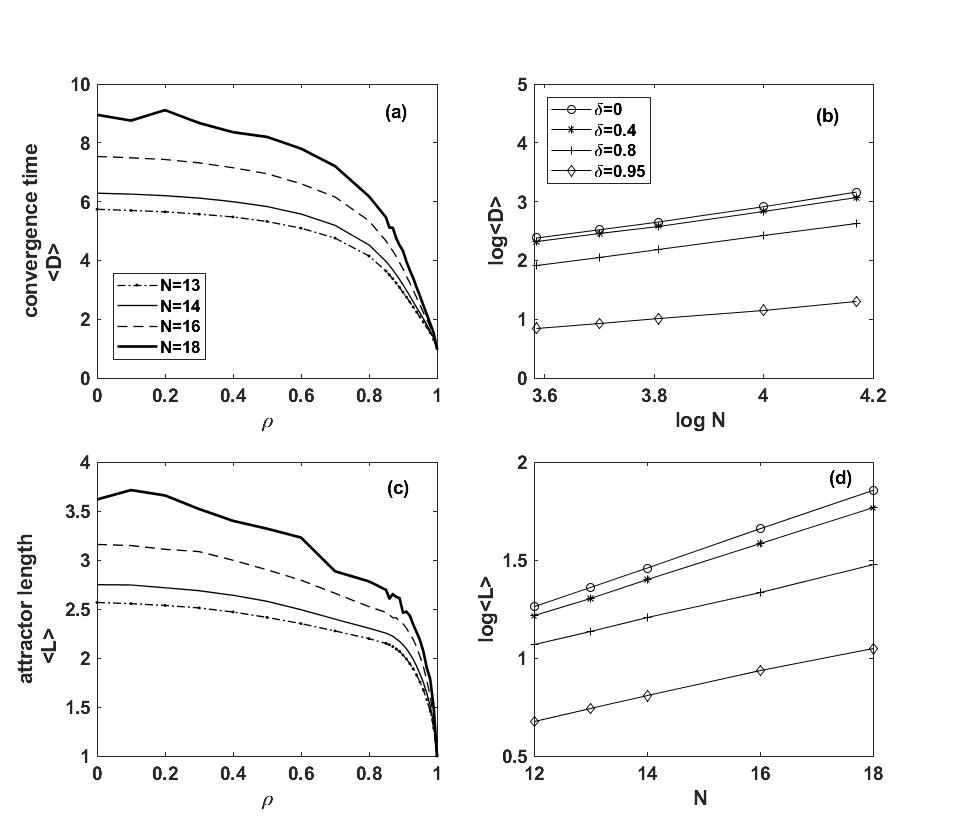}
\caption{The panels on the left, (\textbf{a}) and (\textbf{c}),
show respectively how the average convergence time $\langle D\rangle$  and the
average limit cycle length $\langle L\rangle$ change over different $\rho$ values given $N=13,14 ,16 ,18$.
The panels on the right, (\textbf{b}) and (\textbf{d}), show how  $\langle D\rangle$ and $\langle L\rangle$
scale over different $N$ values given
$\rho= 0,0.4,0.8,0.95$.
{\bf {Increasing sparsity in asymmetric matrices reduces
the convergence time and the attractor lengths.
Thus, sparse matrices recover computation ability
and lose exponentially long limit cycles characteristic of the chaotic regime.}}} \label{fig3}
\end{figure}

A diluted and asymmetric discrete-time recurrent neural network
recovers an optimal storage capacity and stimulus-response association time
compared to a dense asymmetric recurrent neural network.
Greater dilution increases the number attractors and
accelerates the retrieval process. Here, we analyze the same quantities measured in the previous section,
but we consider different dilution values $\rho\in(0,1)$, and keep constant the asymmetry coefficient $\epsilon$=1.
Thus, we consider asymmetric networks at different dilution levels.
In figure~\ref{fig3}, we report how the convergence time $D$ changes with the dilution $\rho$ (\textbf{a}) and how it scales with $N$ (\textbf{b}).
Increasing dilution $\rho$ drastically decreases the average
stimulus-response association
$\langle D \rangle$
for pattern retrieval. In other words, the dilution improves the ability of the network to converge to an attractor.
Figure~\ref{fig3} (\textbf{c}) shows how as dilution increases the length of limit cycles is
reduced.
In highly diluted systems ($\rho$ above $0.7$), a polynomial fit better predicts 
the observed data compared to an exponential fit (\textbf{d}).
The scaling coefficient is $\gamma_{p}$=0.66$ \pm 0.07$ for $\rho$=0.95.
In other words, for diluted networks
the limit cycles length does not scale exponentially with $N$ anymore,
and the ``chaotic" regime ceases.
\begin{figure}[h!]
\includegraphics[width=\columnwidth]{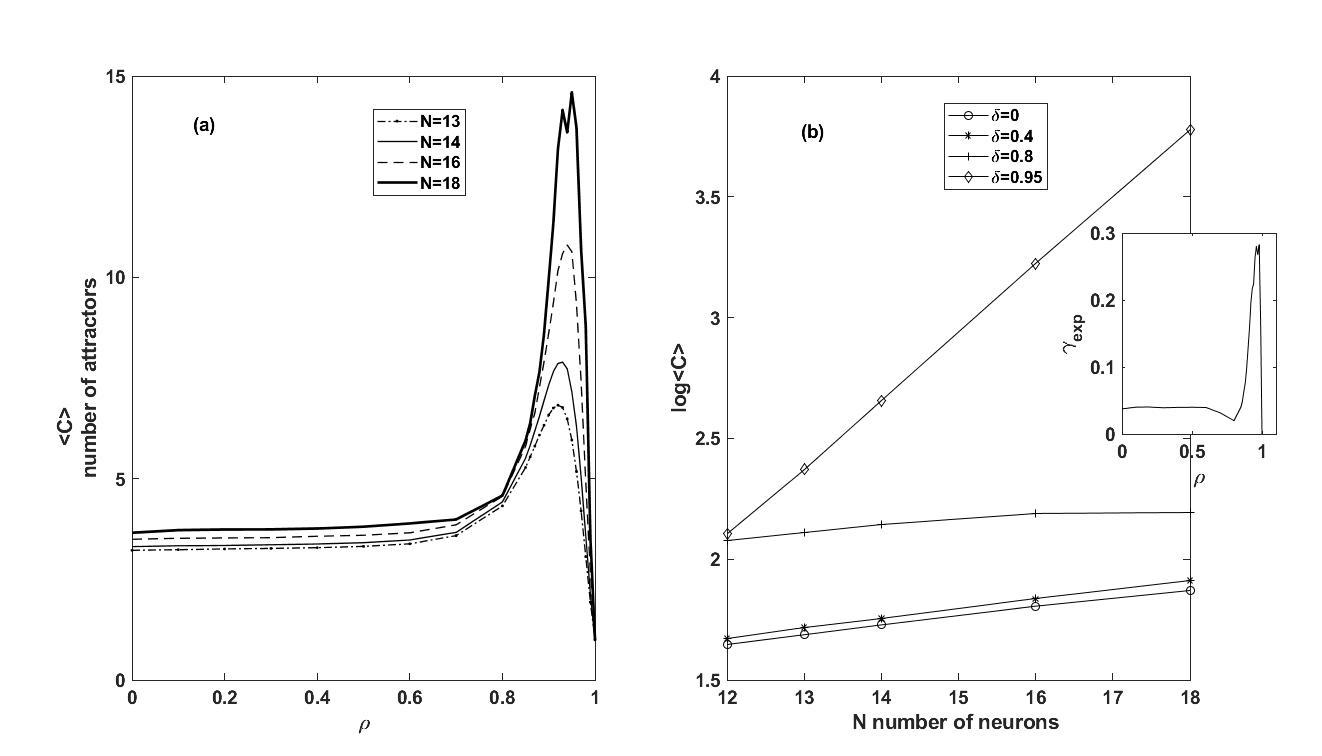}
\caption{(\textbf{a}) The plot shows how $\langle C \rangle$ changes with the dilution coefficient $\rho$ for different values of $N$,
$N=13,14 ,16 ,18$.
(\textbf{b}) The plot shows how $\langle C \rangle$ scales with $N$ for $\rho= 0,0.4,0.8,0.95$.
The inset shows the exponential coefficients for different values of $\rho$.
{\bf {The figure reports the main result of the present work. 
As sparsity in fully asymmetric matrix is increased, an optimal region appears:
the non-monotonic peak in which the network architecture admits an exponentially large number of attractors. Furthermore, as shown in Fig.~\ref{fig3}, in the same regime, the network is able to quickly recall memories converging to the corresponding attractor,
and the attractor is generally a limit cycle with small $L$. 
All these features show an optimal asymmetric/diluted network regime 
that maximizes the storage capacity and the recovery efficiency.
}}} \label{fig4}
\end{figure}

Figure~\ref{fig4} reports the main result of the present work.
It provides direct evidence of the nontrivial effect of dilution:
increasing dilution in an asymmetric network causes the appearance of an unexpected maximum 
for the storage capacity $\langle C \rangle$ at $\rho$=0.95 (\textbf{a}).
This value approximately corresponds to $90\%$ of zeros in the synaptic matrix,  meaning only $10\%$ of the neuron pairs are connected. The height of the peak exponentially scales with the network size $N$ (\textbf{b}) with
$\gamma_{e}$=0.28$\pm0.02$ for $\rho$=0.95,
which is approximately the same scaling value
estimated for $\langle C\rangle$ in fully-connected symmetric networks with $\epsilon$=0, $\rho$=0.

A caveat is due because finite size effects may determine the emergence of the 
peak. 
Indeed, it is unclear, given the current observations, if, for large $N$, the
peak position around $\rho$=0.95 is stable or if it drifts towards $\rho$=1 (finite size effect).
For this purpose, possible future investigation could
be the analytical estimation of the peak position given the thermodynamic limit, which may be 
obtained with the replica method or the more appropriate cavity method \cite{Mezard1987}. 
It is important to point out that if it is actually true that the peak drifts and disappears in the thermodynamic limit,
this implies that its maximum position gets closer and closer to the ``empty" matrix value $\rho$=1.
Given that the peak height will most likely increase exponentially with $N$,
we would find that an exponential number of stimulus-response associations could be stored
in almost-connection-free large networks.

To qualitatively validate the persistence of the peak for large networks,
we implemented a Monte Carlo experiment that estimates the average number of attractors $\langle C \rangle$
for networks with $N=45$, $\epsilon = 1$ and for $0 \le \rho \le 1$. 
For each network, we sampled $2^8$ initial conditions out of $2^{45}$, thus we explore a fraction equal to $10^{-11}$ of the entire initial condition space.
This estimate is biased differently depending on the network's specific values of symmetry $\epsilon$ and dilution $\rho$,
because for certain values of symmetry $\epsilon$ and dilution $\rho$ we have a different distribution of the size of the attraction  basins,
which produces a disproportion in the sampling probability of attractors with small attraction basins. 
For $N=45$, we find that the peak at $\rho=0.95$ is preserved, but that 
the
value of $\langle C \rangle$
relatively raises for $\rho<<0.8$ compared to $\rho>>0.8$, because
for $\rho<<0.8$, $\langle C \rangle$ is overestimated with regards to the estimate at  $\rho>>0.8$.
This indicates the random sampling inapplicability
and the need for an exhaustive scan of the initial condition.
Furthermore, this observation
reinforces the hypothesis that for large $N$ the peak persists around $\rho=0.95$.

\begin{figure}[h!]
\includegraphics[width=\columnwidth]{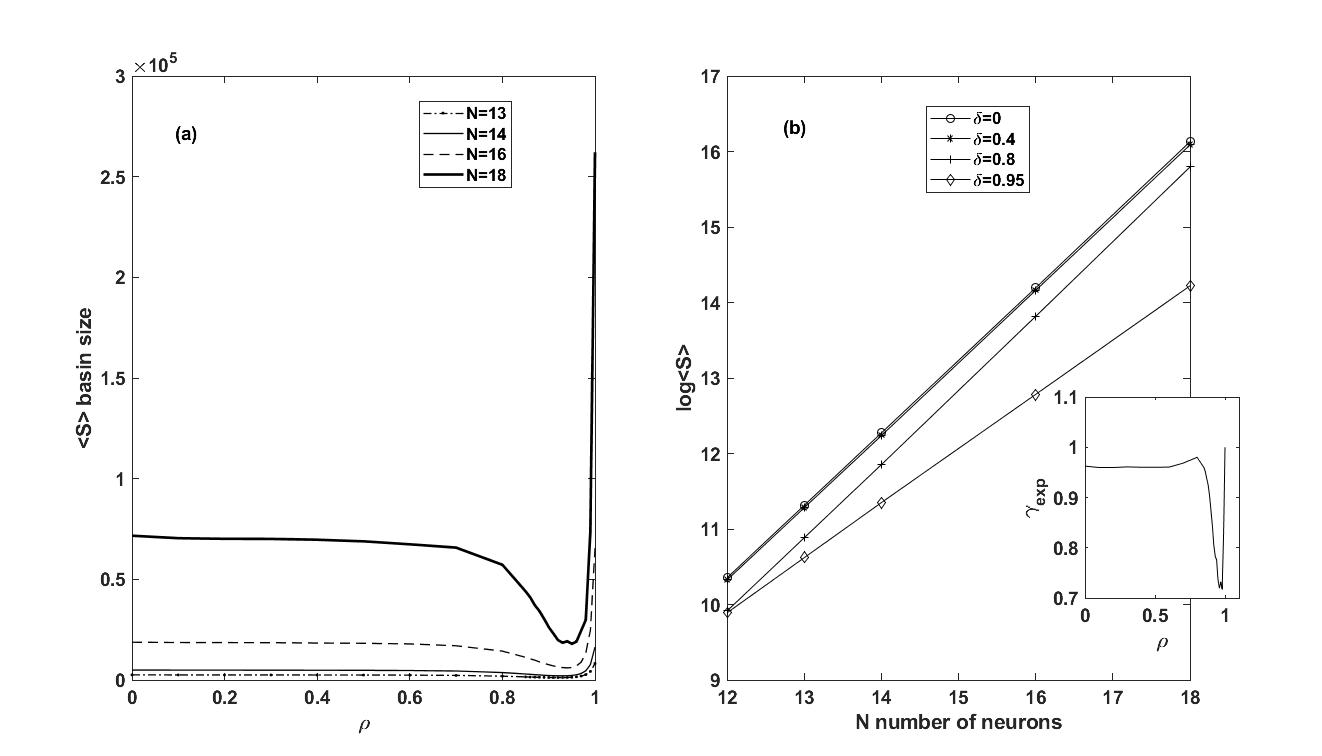}
\caption{
(\textbf{a}) The plot shows how the average basin size $\langle S \rangle$
changes as a function of the dilution parameter $\rho$ for different
values of $N$, $N = 13,14,16,18$. (b) Shows how $\langle S\rangle$ 
scales with $N$ for $\rho=0,0.4,0.7,0.95$. The insert shows
the exponential scaling coefficients for different $\rho$ values.
{\bf {This figure shows that the region with the maximum storage capacity is characterized by many small attraction basins.}}}
\label{fig5} \end{figure}

Figure \ref{fig5} shows how the average basin size $\langle S \rangle$ changes as a function of the dilution and of the network size $N$.
In correspondence of the peak in $\langle C \rangle$, the dimension of the attraction basins decreases (see figure \ref{fig5} (\textbf{a})). The state space tends to be subdivided in a large number of small regions, each with its attractor. 
This and the fast stimulus-response association tells us that the network is able to quickly subdivide the inputs in clusters: in fact, at $\rho$=0.95, the average convergence time is strongly reduced with respect to the chaotic regime and its scaling turns from exponential to polynomial (see figure~\ref{fig3}, (\textbf{a})-(\textbf{b})).
All these observations allow us to conclude that asymmetric-diluted
recurrent neural networks
exhibit optimal information processing
and memory storage for $\rho$=0.95 and $\epsilon$=1.

\section{Discussion}

This paper discusses 
how the dilution and asymmetry
of a discrete-time recurrent neural network connectivity matrix $\mathbf J$ influences the network's
limit behavior storage capacity and its response time.
More precisely, it explores the effect of dilution in a
discrete-time recurrent neural network 
of binary neurons
with an asymmetric excitatory/inhibitory 
connectivity matrix.
Given an arbitrary learning rule,
and an incremental sequence of memories,
a recurrent neural network,
reaches a limit behavior storage limit. We suppose that this limit is
in general characteristic of the particular learning rule. 
To go beyond the limitation of having to assume a particular learning rule,
we consider the 
recurrent neural network limit behavior storage capacity
as the upper bound
of the limit behavior storage capacity of any arbitrary learning rule. 
Furthermore, we assume that the connectivity matrix 
of a recurrent neural network 
which has stored a number of limit behaviors close to the storage limit must 
have a connectivity matrix characterized by a specific structure
with certain values of asymmetry and dilution.
Clearly, this characteristic connectivity structure 
must approach the structure of a recurrent neural network 
with the largest number of attractors
obtained from a pool of all possible networks, because this
should be equivalent to the network that has potentially stored the largest number of memory vectors.
To find this characteristic storing structure,
we sample random connectivity matrices $\mathbf J$ 
of arbitrary asymmetry $\epsilon$ and dilution $\rho$,
and for each matrix we map the full attractor basin for all $2^N$ initial conditions.
This allows us to examine the neural network's storage capacity,
and measure how quickly it clusters $2^N$ initial conditions into the corresponding attractors. 
We considered the scenario in which the network is potentially able to store memories both as fixed point attractors
and limit cycles attractors, without imposing any a priori learning role.

We found two regions in the asymmetric/diluted space $(\epsilon,\rho)$ of all possible sampled networks
in which neural networks exhibit optimal storage capacity.
The first optimal storage capacity region contains 
asymmetric $\epsilon =1$
and diluted $\rho \sim 0.95$ connectivity matrices. This means that a large fraction ($\sim90\%$)
of elements in the connectivity matrix are zero.
The second optimal region incorporates fully-symmetric $\epsilon =0$,
and fully connected matrices $\rho=0$, in this region almost all connectivity matrix elements are
non-null.
Similar connectivity as in the first region is observed in
the neocortex and in the
CA3 region of the hippocampus \citep{Witter2010,Perin2011}.
These natural neural networks 
are implicated in memory storage and retrial.
These observations are coherent with previous analytical and computational observations \citep{Kim2017,vanVreeswijk1998, Monteforte2012}, 
but it was unknown how this optimal connectivity changes in the surrounding of 
the optimal value of $\rho$ and was considered
as substantial evidence against Hebbian learning.
For this reason Hebbian learning is considered as a poor model of learning in animal neural networks. 
Contrarily, the second region is only predicted by Hebbian learning but is not
observed in natural neural networks.
Furthermore, we found that in the fully-connected $\rho=0$ and fully-asymmetric region $\epsilon =1$,
the recurrent neural networks exhibit a very low storage capacity, and large basins of attraction, which implies that the
network is no longer able to distinguish and separate different external stimuli. In addition,
fully-connected and asymmetric regions have very long (glassy) recognition times, which compromise the ability
of the recurrent neural network to respond to external stimuli.
From a neurobiological prospective, this means that when a recurrent neural network drifts out of its optimal state
for some external cause such as a disease,
then the network becomes less effective in separating different stimuli, and discriminating errors from signals.
In addition, its response times become longer.
In line with these observations, \cite{Tang2014} report that the brain of patients affected by
autism spectrum disorders (ASDs) presents an altered connectivity, or dilution, compared
to healthy individuals, and specifically reports an alteration in the neocortex connectivity.
Suppose that the brain regions involved in ASDs are displaced
from their optimal asymmetric/diluted region,
we could argue, given our observations, that this displacement may cause a disruption of
the ability to separate different stimuli in distinct responses, and produce longer response times.

\section{Conclusion}
From our exploration of the role of connectivity and symmetry in recursive neural networks,
we find two regions that optimize limit behavior storage and signal-response association. The first
region is composed of asymmetric/diluted networks, and the second region is formed of symmetric/fully-connected networks.
Furthermore, we found a third region made of asymmetric/fully-connected networks characterized by chaotic and glassy limit behaviors.
From these results we are left with the question 
of why adaptation and evolution selected the first region. Is this because
more non-zero elements in the connectivity matrix corresponds to more costly connections?
Is it because fully-connected networks imply the existence of two axons between any two neurons and this would be
spatially impossible if not technically implausible
for large network sizes $N$?
Is it because the natural learning rules 
that guide neural networks development force them to dynamically evolve in the second region?

Lastly, to partly overcome the smallness of $N$, we have also analyzed the scaling properties of the main measured quantities. 
We have found that the scaling behavior for the averaged number of length 1 cycles in fully-connected symmetric networks
are perfectly in agreement with theoretical values found by \cite{Tanaka1980}. Thus the averaged number of length 1 cycles
is not biased by the finite size effects. In addition, in the analyzed range for small $N$, all the scaling laws appear highly robust.
These observations do not guarantee that the same occurs for the
scaling laws of other observable quantities in diluted networks. Nevertheless, it gives us confidence
in the extrapolation of the scaling laws observed in this paper.

\section*{Conflict of Interest Statement}
The authors declare that the research was conducted in the absence of any commercial or financial relationships that could be construed as a potential conflict of interest.

\end{document}